\documentclass[journal=cmatex,manuscript=article]{achemso}

\usepackage[version=3]{mhchem} 
\usepackage{subcaption}
\usepackage{standalone}
\usepackage{tikz}
\usepackage{helvet}
\usetikzlibrary{positioning}

\graphicspath{figures}

\newcommand{\pp}[0]{$P^2$}
\newcommand{\siref}[1]{{Supporting Fig. S#1}}
\newcommand{\fref}[1]{Fig. \ref{#1}}
\newcommand{\fcc}[0]{$Fm\overline{3}m$}
\newcommand{\bcc}[0]{$Im\overline{3}m$}
\newcommand{\hcp}[0]{$P6_3/mmc$}
\newcommand{\dnn}[0]{$d^{NN}_{H-H}$}
\newcommand{\qh}[0]{$\Delta Q_H$}

\graphicspath{{figures/}}
\author{Mgcini Keith Phuthi}
\affiliation{Department of Mechanical Engineering, University of Michigan, Ann Arbor, Michigan 48103, USA}
\alsoaffiliation{Department of Mechanical Engineering, Carnegie Mellon University, Pittsburgh, Pennsylvania 15213, USA}
\author{Pin-Wen Guan}
\affiliation{Department of Mechanical Engineering, Carnegie Mellon University, Pittsburgh, Pennsylvania 15213, USA}
\author{Russell J. Hemley}
\affiliation{Departments of Physics, Chemistry, and Earth and Environmental Sciences, University of Illinois Chicago, Chicago, Illinois 60607, USA}
\author{Venkatasubramanian Viswanathan}
\affiliation{Department of Aerospace Engineering, University of Michigan, Ann Arbor, Michigan 48103, USA}
\alsoaffiliation{Department of Mechanical Engineering, Carnegie Mellon University, Pittsburgh, Pennsylvania 15213, USA}
\email{venkvis@umich.edu}

\title[An \textsf{achemso} demo]
  {Stability and Structure of Binary Metal Hydrides under Pressure, Electrochemical Potential and Combined Pressure-Electrochemistry}

\abbreviations{DAC,P2,HER,RHE,SHE}
\keywords{Metal Hydride, Electrochemistry, Phase Diagrams}

\begin{document}








\begin{abstract}
  Metal hydrides can be tuned to have a diverse range of properties and find applications in hydrogen storage and superconductivity. Finding methods to control the synthesis of hydrides can open up new pathways to unlock novel hydride compounds with desired properties. We introduced the idea of utilizing electrochemistry as an additional tuning knob and in this work, we study the synthesis of binary metal hydrides using high pressure, electrochemistry and combined pressure-electrochemistry. Using density functional theory calculations, we predict the phase diagrams of selected transition metal hydrides under combined pressure and electrochemical conditions and demonstrate that the approach agrees well with experimental observations for most phases. We use the phase diagrams to determine trends in the stability of binary metal hydrides of scandium, yttrium and lanthanum as well as discuss the hydrogen-metal charge transfer at different pressures. Furthermore, we predict a diverse range of vanadium and chromium hydrides that could potentially be synthesized using pressure electrochemistry. These predictions highlight the value of exploring pressure-electrochemistry as a pathway to novel hydride synthesis.
\end{abstract}

\section{Introduction}
Metal hydrides have gained significant interest in the last few decades for applications in hydrogen storage \cite{schneemann_nanostructured_2018, rivard_hydrogen_2019, el_kharbachi_metal_2020}, as control rods for nuclear reactors \cite{bannenberg_optical_2019}, switchable mirrors \cite{nagengast_epitaxial_1999}, catalysis \cite{cao_vanadium_2021,jackson_preparation_2008,yan_dinitrogen_2022}. More recently, there has been a huge surge in interest due to experimentally measured superconductivity in ``superhydride" materials \cite{geballe_synthesis_2018,somayazulu_evidence_2019}. Superhydrides, sometimes referred to as polyhydrides, are materials with ``unusually high hydrogen-to-metal ratios". Superhydrides formally have a stoichiometric ratio of 6 hydrogen atoms or more relative to the other elements of the compound\cite{geballe_synthesis_2018}.

The challenge in synthesizing superhydrides lies in stabilizing the hydrogen in the structure as it typically escapes as hydrogen gas. To stabilize a hydride therefore requires conditions that favor formation of the hydride over the formation of hydrogen gas. One way to stabilize hydrides is to apply external pressure until the chemical potential of the hydrogen in the \ce{H2} gas is lower in the solid hydride whose energetics are less affected by the external pressure \cite{wipf_metal-hydride_1997}. This approach has seen significant success in stabilizing superconducting hydrides with near room-temperature critical temperature ($T_c$) values. Examples of experimentally verified superconductors include \ce{LaH10} ($T_c=260$ K at 180-200 GPa)\cite{somayazulu_evidence_2019, drozdov_superconductivity_2019}, \ce{H3S} ($T_c=203$K at 141 GPa) \cite{drozdov_conventional_2015} and many others.

At low pressures below 0.1 MPa, a sample of the pure metal can be exposed to pressurized hydrogen gas in a pressure cell, usually to form hydrides with ratios less than 3 \cite{yukawa_compositional_2003}. In contrast, the pressure regimes necessary for synthesizing superhydrides are typically extreme, on the order of $\sim$100 GPa, only realizable in diamond anvil cells (DACs). For these studies, DAC sample sizes are in the range of 100 $\mu$m down to 20 $\mu$m in diameter\cite{obannon_contributed_2018}. The technological and engineering challenges that need to be surmounted to reliably synthesize and use these materials in applications are therefore quite significant.

A different approach employs electrochemical loading of hydrogen, typically used in hydrogen storage applications. Hydrogen storage is desirable because it can achieve a higher energy density for longer lifetimes than lithium ion batteries with lower environmental impact \cite{el_kharbachi_metal_2020}. The advantage of electrochemical hydrogen loading as opposed to pressurization of \ce{H2} gas and other methods is that the hydrogen can be stored either as ions, or as physisorbed/absorbed atoms or molecules, making the storage much safer. It is also easier to monitor hydrogen levels in a materials using voltammetry, amperommetry, electronic impedance spectroscopy (EIS) or various other methods \cite{kaur_review_2019}. There are still a number of challenges related to making practical large-scale hydrogen storage devices. It is therefore critically important to understand which materials and under which conditions these materials take up hydrogen as well as the ease with which hydrogen can be absorbed and desorbed by the material. The ideal material loads hydrogen quickly under typical storage conditions, near ambient conditions and releases hydrogen at high rates \cite{wipf_metal-hydride_1997,el_kharbachi_metal_2020}.

Techniques that combine these approaches have been of previous interest but typically only at megapascal pressures\cite{giovanelli_electrochemistry_2004}. In an earlier work\cite{guan_combining_2021,guan_low-pressure_2022}, we proposed the combined use of pressure and electrochemistry, termed the \pp{} approach to stabilizing hydrides, including superhydrides at gigapascal pressures. We build upon this work by considering different metal hydrides while also giving improved and more detailed phase diagrams. In the methods section, we discuss the theory behind pressure-potential (\pp{}) phase diagrams as well as the computational details and considerations made in the calculations. We then present \pp{} phase diagrams of scandium, yttrium, lanthanum as well as vanadium and chromium before finally discussing the results.

\section{Methods}
\subsection{Pressure-Potential Hydride Phase Diagrams}\label{sec:p2}

\begin{figure*}[t!]
    \centering
    \begin{subfigure}[t]{0.45\linewidth}
        \centering
        \includegraphics[width=1\textwidth]{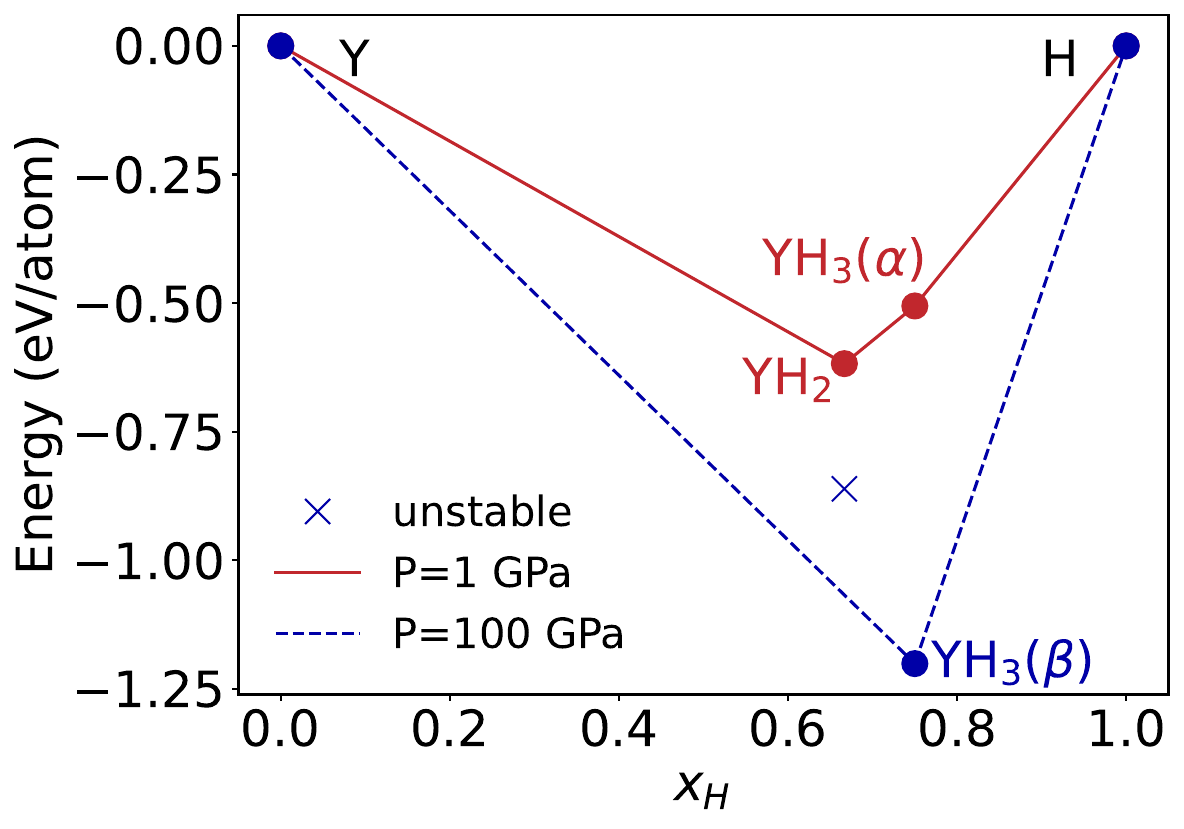}
        \caption{}\label{fig:peffect}
    \end{subfigure}
    \begin{subfigure}[t]{0.45\linewidth}
        \centering
        \includegraphics[width=\textwidth]{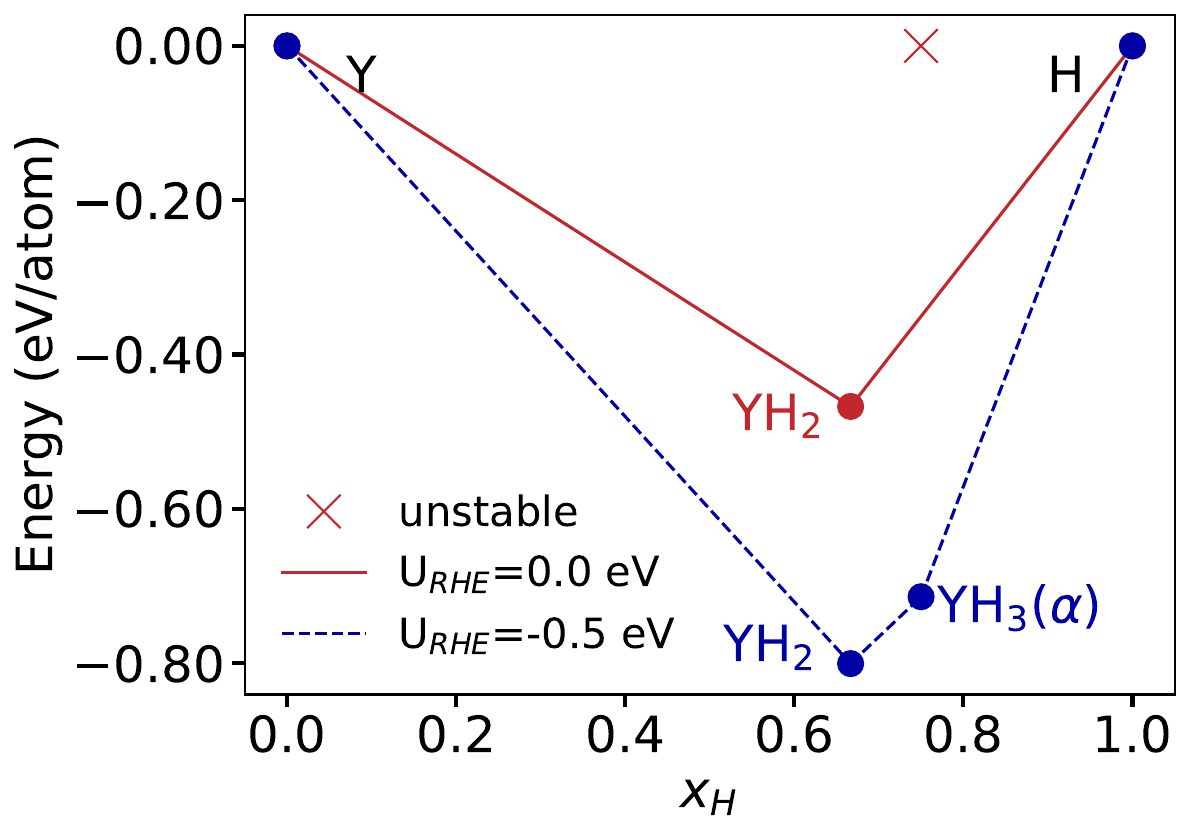}
        \caption{}\label{fig:ueffect}
    \end{subfigure}\\
    \caption{Demonstrating the effect of a) pressure and b) electrocehmical potential on the stability of yttrium hydrides using the convex hull approach. Varying the conditions determines which phases appear on the convex hull and are therefore stable.}
\end{figure*}

First, we consider the loading of hydrogen using only pressure which can be summarized by the equation
\begin{equation}
\ce{M(s) + xH -> MH_x(s)}.
\end{equation}
We consider the hydrogen source to be a reservoir that could contain hydrogen in any state, not just gaseous form, as it is fluid or solid form at higher pressures. Since we are considering the case of fixed temperature ($T$) and pressure ($P$), the Gibbs free energies of each component are the relevant quantities by which to determine the stability of different phases i.e. $G_M(P,T)$, $G_H(P,T)$ and $G_{MH_x}(P,T)$. The convex hull concept, discussed extensively in the literature and depicted in \fref{fig:peffect} is a conventional approach for determining material stability given free energies\cite{bartel_review_2022}. A phase or crystal structure is said to be stable if the relevant free energy, referenced to the pure elements at the same $P,T$, and therefore called the formation energy, is negative enough that it appears on the convex hull. 

The variation of the free energy is different for different crystal structures, therefore the structures that appear at different pressures could be different, even for the same composition.  We demonstrate this in \fref{fig:peffect} where we show the convex hulls at two different pressures. At 1 GPa, both \ce{YH2} and \ce{YH3} appear on the hull and are therefore stable. When pressure is increased to 100 GPa, two changes are observed in this case. First, the \ce{YH2} becomes unstable as it ceases to appear on the convex hull. Secondly, the \ce{YH3} ($\alpha$) phase with \hcp{} symmetry becomes less stable than the \ce{YH3} ($\beta$) phase with \fcc{} symmetry but the composition \ce{YH3} remains on the convex hull. With increasing pressure, due to the more rapid increase in the hydrogen chemical potential relative to the metal under pressure, higher hydrides will continue to be stabilized and appear on the convex hull. This is how superhydrides are stabilized in DAC experiments.

In the electrochemical case, when an electrolyte with an excess of \ce{H+} (or \ce{H-}) ions is used in an electrochemical cell, it is possible to ``load" hydrogen into the electrode by passing a current due to a potential difference ($U$) between two electrodes. The process can be summarized in the following equation if the anode is say, a metal \ce{M}, and a negative potential is applied.
\begin{equation}
\ce{M(s) + xH+ + xe- -> MH_x(s)}.
\end{equation}
The change in Gibbs free energy, $\Delta G$ for this reaction under given conditions of pressure, temperature and applied electrochemical potential is given by
\begin{equation}\label{eq:grhe}
    \Delta G = G_{MH_x}(P,T) - G_M(P,T) - xG_{H}(P,T)+xeU_{\text{RHE}}.
\end{equation}
\cite{guan_combining_2021}$G_{MH_x}$ and $G_M$ are the free energy of the hydride and pure metal respectively and are approximated with Density Functional Theory at $T=0$K and pressure $P$.  The further approximation of negligible entropy is used to simplify the calculations such that only the enthalpy ($H$) is considered i.e. $G\approx H_{\text{DFT}}(P,T=0) = E_{DFT}+PV$ where V is the unit cell volume as is commonly assumed for solids. $E_{DFT}$ is the DFT energy of the crystal structure with the lowest energy for the given composition with ionic positions optimized so that forces on all atoms less than $0.01$eV/\AA{}. As shown in \fref{fig:peffect}, this could be different structures at different pressures. Candidate structures are obtained from crystal structure searches in the literature, the Materials Project \cite{jain_commentary_2013} and by substituting the metallic species in known crystal structures.

The hydrogen free energy $G_H(P,T=300 K)$ is challenging to calculate consistently as hydrogen exists as gas, liquid and solid throughout the pressure range considered. In this work, we approximate $G_H$ using the following piecewise function

\[ G_H(P,T=300 K) \approx \begin{cases} 
      E_{DFT}^{\ce{H2}}(P=0,T=0) +f_{\text{trans}}(T) + f_{\text{rot}}(T) + \delta_{\text{expt}-\text{gas}}& P < 0.004 \\
      g_{\text{expt.}}(P,T=300) + \delta_{\text{expt}-\text{solid}} & 0.004\leq P\leq 10 \\
      H_{DFT}(P,T=0) & P > 10
   \end{cases}
\]

\noindent where the pressures are specified in gigapascal units and temperatures in Kelvins. $E_{DFT}^{\ce{H2}}$ is the energy per atom of the DFT relaxed hydrogen molecule at zero pressure, $f_{\text{trans}}$ and $f_{\text{rot}}$ are the translational and rotational free energies per atom in the ideal gas approximation. The vibrational degrees of freedom  are not considered as they are frozen out below 1000 K\cite{minor_greiner_2014}. $g_{\text{expt.}}$ is the Gibbs free energy per atom of hydrogen from experimental measurements available in the NIST database \cite{linstrom_nist_2001}. $\delta_{\text{expt}-\text{gas}}$ and $\delta_{\text{expt}-\text{solid}}$ are the shifts in energy required to align the energies where the piecewise function connects. We chose to reference against the high pressure solid hydrogen as it is more consistent with the rest of the calculated energy values. While this approximation may seem rather crude, we show that it gives sufficiently accurate predictions in the results section.

The potential ($U_{\text{RHE}}$) is referenced to the reversible hydrogen electrode (RHE) to avoid considering the pH of the electrolyte, which would have to be accounted for if the standard hydrogen electrode (SHE) scale was used. The effect of increasingly negative $U_{\text{RHE}}$ is shown in \fref{fig:ueffect}. Since the effect of $U_{\text{RHE}}$ carries a multiplier of $x$, higher hydrides with $x>1$ will tend to be more stabilized by the potential than lower hydrides until they are eventually more stable as long as $G_H$ does not increase more rapidly than the hydride.


A challenge with electrochemical loading is the competition with the hydrogen evolution reaction (HER) which causes hydrogen to leave the hydride at pressures where \ce{H2} is stable by the reaction
\begin{equation}
    \ce{H+_{(\text{aq})} + e- <=> \frac{1}{2}H2_{(\text{g})}}.
\end{equation}
In the SHE reference, the above HER is considered to be in equilibrium. A more negative applied potential, which also corresponds to a more negative potential on the RHE scale therefore stimulates the forward reaction in competition with hydride formation. Many metals are also catalysts for the HER \cite{li_recent_2020}. For each metal, there is an HER onset potential at which all the steps in the catalytic reaction become downhill and HER is spontaneous and therefore dominates. Some of the metals considered in this work have HER onset potentials calculated by Guan et al.\cite{guan_combining_2021}. Below the HER onset potential, one can expect hydrogen gas to bubble off making it challenging to form the hydride, therefore we consider this a limit of synthesizability. We will refer to this limit as the ``HER line" below which hydrogen evolution is expected to dominate. It is, however, only a guide; experimentally, it is in fact advantageous to use a large negative potential even if hydrogen is bubbling unless the bubbling damages the electrode\cite{berlinguette_revisiting_2019}. Hydrogen gas also tends to bubble at potentials less negative than the HER onset potential anyway, especially if the metal is coated with for example palladium to protect it from oxidation and help with hydrogen absorption \cite{benck_producing_2019}.

The advantage of applying pressure in this electrochemical setting is that it opposes the HER, making it more favorable for the hydride to form. The \pp{} approach therefore takes advantage of the very large driving forces achievable using an easily achievable potential while using the pressure as a HER suppression and stabilization mechanism. However, the stability of the hydride formed once the applied pressure or potential is removed however depends on the kinetics of the decomposition process.

Applying Eq. \eqref{eq:grhe}, we can approximate the formation energies for different hydrides and predict the how their stability changes with $P$ and $U_{\text{RHE}}$ to generate \pp{} phase diagrams. Because we predict which crystal structure is most stable at different conditions, we also show this in the phase diagram. This is important as material properties depend heavily on the crystal structure of the material, not just the composition.

\subsection{DFT parameters}
The DFT parameters chosen were such that the Brillouin zone sampling density, plane wave and density cutoffs were converged to $<$1meV/atom to ensure consistency. DFT calculations were done using Quantum Espresso \cite{giannozzi_quantum_2009} within the Generalized Gradient Approximation using the Perdew-Burke-Eizenhoff exchange correlation functional \cite{perdew_generalized_1997}. Pseudopotentials used were obtained from the standard solid-state pseudopotentials (SSSP) website \cite{prandini_precision_2018}. A uniform Brillouin Zone spacing of 0.02\AA$^{-1}$ with a Monkhorst-Pack \cite{monkhorst_special_1976} sampling procedure was used. To help with convergence of the the Fermi surface, Methfessel-Paxton \cite{methfessel_high-precision_1989} smearing using a smearing width of 0.27 eV was chosen for metals. Pressures in the unit cell optimizations were converged to 0.5 GPa. We found these parameters to be suitable for hydrides using convergence tests. The high-throughput workflow used for DFT calculations and analysis was built using the Atomic Simulation Tools package (ASIMTools) \cite{phuthi_asimtools_2024}.

\subsection{Charge and Structure Analysis}
Bader charge analysis was performed using the Critic2 software\cite{otero-de-la-roza_critic2_2014}. Only the valence electrons (pseudodensities) were considered in the charge analysis as we are only interested in changes in the charge distributions. This provides a balance between meaningful analysis and computational cost. The integration of charges around basins of attractors was performed using the Yu-Trinkle method.\cite{yu_accurate_2011}. Manipulation of atomic structures was performed using the Atomic Simulation Environment \cite{larsen_atomic_2017} and PyMatgen \cite{ong_python_2013} tools and visualizations were performed using Ovito \cite{stukowski_visualization_2009}.

\section{Results and Discussion} \label{sec:results}

We present \pp{} phase diagrams for rare earth metal hydrides in group 3 of the periodic table which have been of significant interest for high temperature superconductivity and highlight group trends. Furthermore we consider metal hydrides of early transition metals in period 4 for which extensive structure search data is available. The candidate structures considered largely came from previous crystal structure searches available in the literature \cite{ye_high_2018,liu_potential_2017,li_superconductivity_2017,yu_pressure-driven_2015}. These structures were further supplemented with structures from the Materials Project \cite{jain_commentary_2013} and for scandium, yttrium and lanthanum substitutions of the metal ions were included.

\subsection{Rare Earth Metal Hydrides}

\begin{figure*}[t!]
    \centering
    \begin{subfigure}[t]{0.6\textwidth}
        \centering
        \includegraphics[height=6cm]{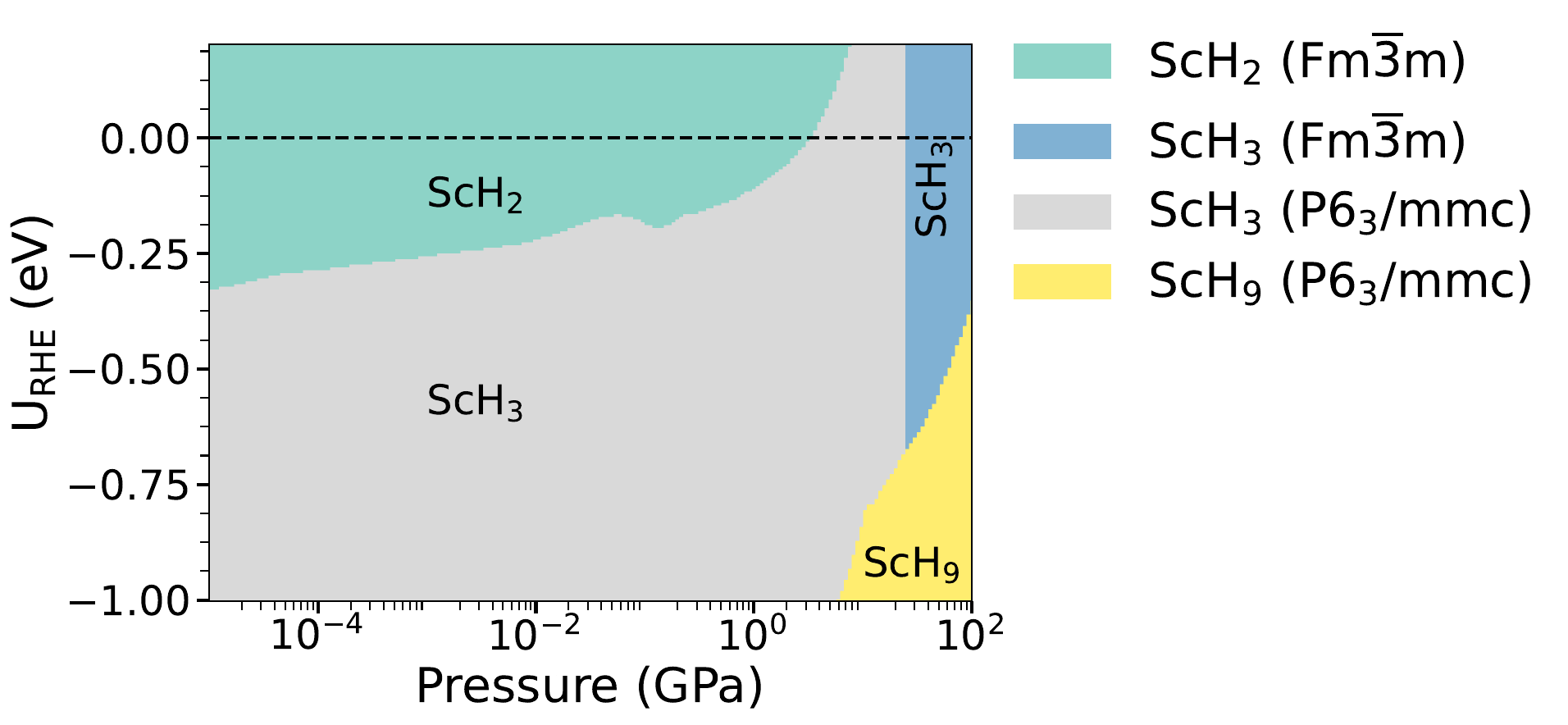}
        \caption{}\label{fig:sch}
    \end{subfigure}\\
    \begin{subfigure}[t]{0.6\textwidth}
        \centering
        \includegraphics[height=6cm]{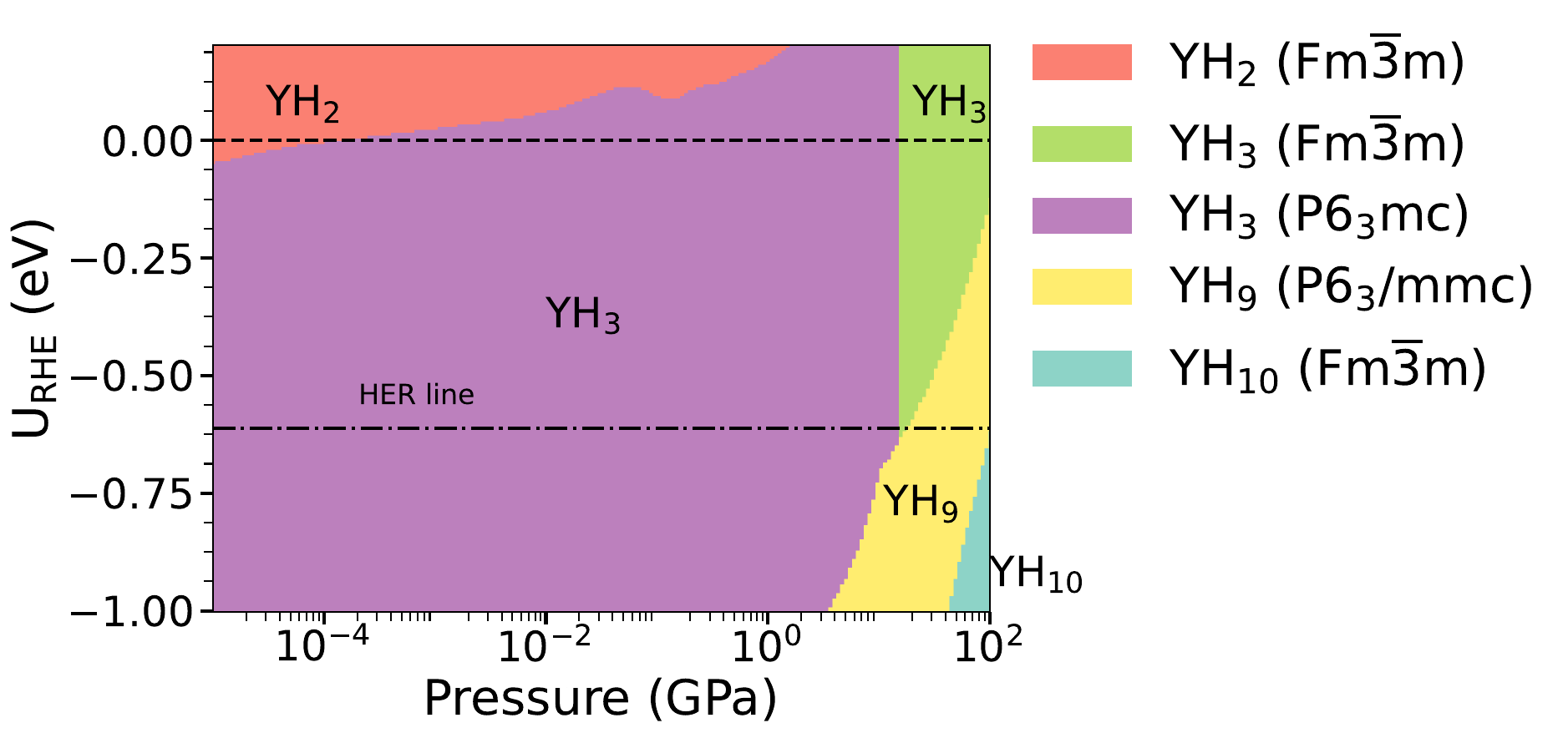}
        \caption{}\label{fig:yh}
    \end{subfigure}\\
    \begin{subfigure}[t]{0.6\textwidth}
        \centering
        \includegraphics[height=6cm]{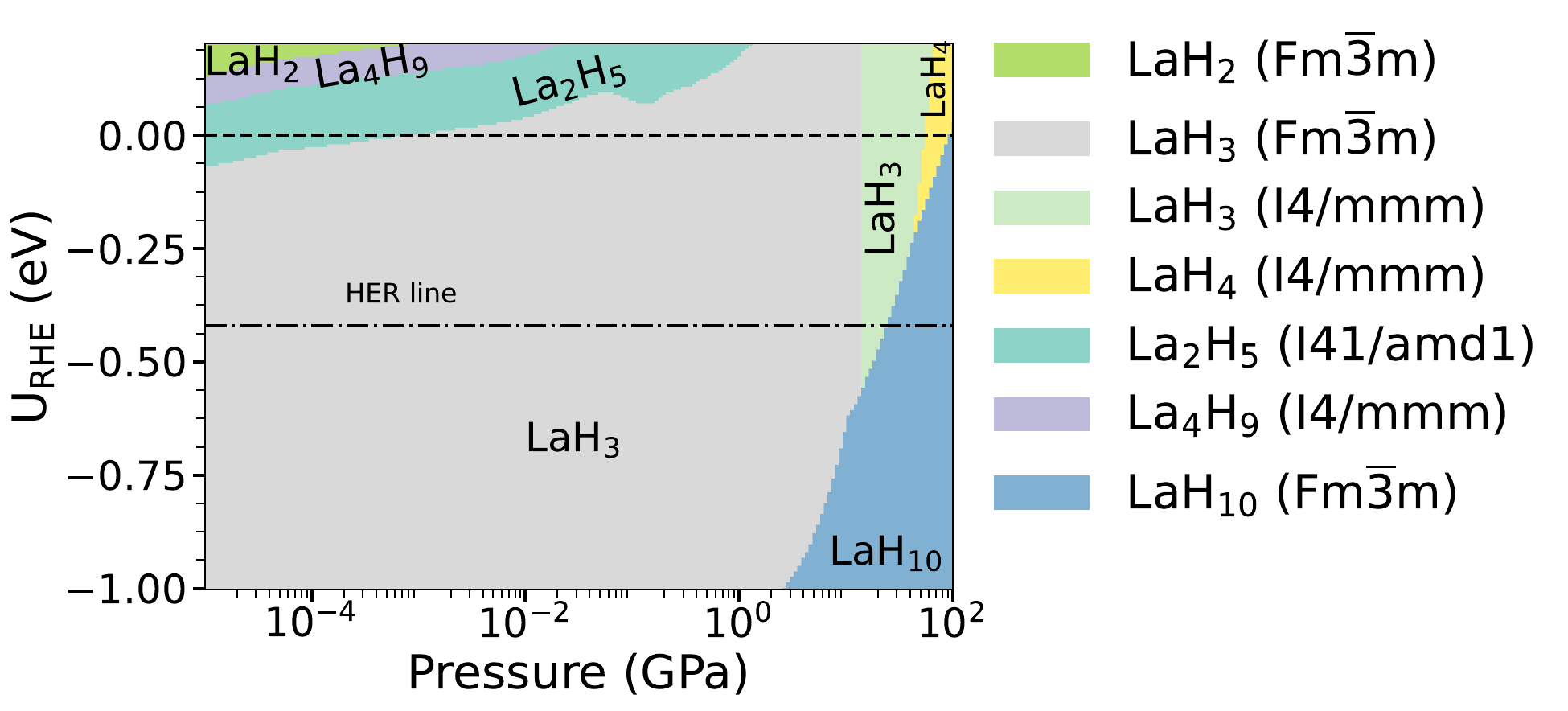}
        \caption{}\label{fig:lah}
    \end{subfigure}
    \caption{\pp{} phase diagrams for a) scandium b) yttrium and c) lanthanum hydrides. The dashed line is a guide to the eye for $U_{\text{RHE}}=0$. The dash-dotted line shows the potential below which HER is expected to dominate as calculated by Guan et al.\cite{guan_combining_2021}}
\end{figure*}

Hydrides of the rare-earth elements have been of particular interest as they have been shown to have high superconducting critical temperatures at extreme pressures\cite{drozdov_superconductivity_2019,kong_superconductivity_2021}. Guan et al. predicted the \pp{} diagrams for yttrium and lanthanum. We update these phase diagrams, including scandium hydrides, accounting for more crystal structures.

In \fref{fig:sch}, we present the scandium hydride \pp{} diagram for the first time. Experimentally, the pure scandium $\alpha$ phase is known to have \hcp{} symmetry and can readily dissolve up to about 50 atomic percent hydrogen at moderate pressures below 0.13 MPa before transitioning to a $\delta$ phase with \fcc{} symmetry with formula \ce{ScH2} when hydrogen sites are fully occupied \cite{manchester_ce-h_1997}. We predict that this \ce{ScH2} (\fcc{}) structure is stable at ambient pressure and under no electrochemical potential.
On the application of pressure, the \ce{ScH3} (\hcp{}) phase is predicted to form as with yttrium hydrides but has not been observed for scandium in experiments. However, Shao et al.\cite{shao_superconducting_2021} stabilized \ce{ScH3} (\fcc{}) against \ce{Sc} and \ce{ScH} above 130 GPa appearing in the \pp{} diagram above 20 GPa. They also performed experiments to show that \ce{ScH3} (\fcc{}) has a superconducting critical temperature of 18.5 K at 131 GPa \cite{shao_superconducting_2021}. The synthesis of either \ce{ScH3} phase could potentially benefit from the \pp{} approach as more moderate conditions of pressure or potential are required when combining the two approaches, but perhaps not significantly for the superconducting \ce{ScH3} (\fcc{}) phase. We predict that \ce{ScH2} would take up more hydrogen to form \ce{ScH3} (\hcp{}) under the application of a relatively low potential of $U_{\text{RHE}}$=-0.33V at ambient pressure. However, few electrochemical loading experiments have been performed for scandium. Ye et al. performed a comprehensive computational crystal structure search and discussed the observed phases under different pressures and their superconducting properties. In addition to the structures shown in \fref{fig:sch}, they found \ce{ScH4}, \ce{ScH6} and others to be stable at higher pressures not considered here \cite{ye_high_2018}.

Yttrium is expected to have a phase diagram similar to that of scandium, since it is in the same group of the periodic table and thus similar electronic structure. Yttrium hydrides have been considered for their very high superconducting critical temperatures up to 227 K and 243 K for \ce{YH6} and \ce{YH9} respectively at extreme pressures above 200 GPa\cite{kong_superconductivity_2021, snider_synthesis_2021}. \ce{YH3} has also been considered for uses in switchable mirrors\cite{nagengast_epitaxial_1999}.

A refined \pp{} diagram for the yttrium hydride system building on previous work by Guan et al.\cite{guan_combining_2021} is shown in \fref{fig:yh}. There are two major differences to note between the phase diagram presented in this work and that by Guan et al. The first difference is that there is no YH9 at low pressure. In the work by Guan et al.\cite{guan_combining_2021}, all the structures from crystal structure prediction were considered as valid candidates. In this work, we exclude the structures with a broken metal lattice despite having lower enthalpy. A more robust prediction would be possible by the inclusion of entropic and nuclear quantum effects as previously considered for lanthanum super hydrides \cite{kaneko_hydrogen_2011,liu_dynamics_2018}, but is beyond the scope of this work. The second key difference is that we have added structural information to the phase diagram and we note any pressure-driven phase transitions in the phase diagram which were accounted for but not explicitly plotted in previous work. These phase transitions can significantly affect the boundaries between different compositions on both axes and are critical to identifying if the stable hydride has desirable properties such as superconductivity. 

The phases at $U_{\text{RHE}}=0$, have been experimentally observed in DAC experiments \cite{}. The \ce{YH2} phase is experimentally known to transform to \ce{YH3} (\hcp{}) phase then to a \ce{YH3} (\fcc{}) phase near 10 GPa\cite{kong_superconductivity_2021}. We show the phase transition schematically in \fref{fig:peffect}. This \fcc{} phase is followed by \ce{YH4}, \ce{YH6} and \ce{YH9} with increasing pressure as shown in \siref{1}. Interestingly, while \ce{YH9} has only been synthesized at $\sim$100 GPa pressures, the \pp{} diagram suggests that a \pp{} approach lowers the synthesis pressure by an order of magnitude. However, the significant practical considerations in performing such an experiment at $\sim$10 GPa pressure would have to be overcome. The phases at ambient pressure and above the HER line, namely \ce{YH2} and \ce{YH3} (\hcp{}), have been previously synthesized in electrochemical experiments with yttrium hydride thin films \cite{molten_optical_1996,kooij_situ_1999}. In these experiments, the predicted change in structure as well as changes in electronic and increasing optical transparency are observed as the hydrogen ratio increases from 2 to 3. The yttrium \pp{} diagram aligns very well with experiment, due to the depth and variety of theoretical, electrochemical and high-pressure experiments that have been performed for the system. 

Lanthanum superhydride was the first metal hydride to exhibit near room temperature superconductivity with $T_c$=260 K for \ce{LaH10} \cite{somayazulu_evidence_2019}. At ambient pressures, lanthanum hydrides have been considered for nitrogen fixation \cite{yan_dinitrogen_2022} and as candidates for fast hydride ion (\ce{H-}) conduction \cite{izumi_electropositive_2023}, a phenomenon which has also been predicted using ab initio molecular dynamics in compressed \ce{LaH10}\cite{liu_dynamics_2018}. The predicted \pp{} phase diagram is shown in \fref{fig:lah}. Lanthanum has perhaps the highest number of experimentally observed binary metal hydrides \cite{laniel_high-pressure_2022}. Despite significant interest in synthesizing lanthanum hydrides by pressurization, there are few studies synthesizing lanthanum hydrides electrochemically. Pure lanthanum metal takes the HCP (\hcp{}) structure in the $\alpha$ phase and this structure can dissolve hydrogen in small amounts before transitioning to a stoichimetric hydride \ce{LaH3} without an \ce{LaH2} phase like scandium and yttrium. \ce{LaH3} at low pressure has \fcc{} structure and is an insulator that becomes a semimetal/metal above 100 GPa \cite{geballe_synthesis_2018,drozdov_superconductivity_2019} unlike  the \hcp{} structure predicted in scandium and yttrium systems. \ce{LaH3} readily forms in the presence of hydrogen, agreeing with the \pp{} prediction. We do however predict the existence of \ce{La2H5} and potentially \ce{La4H9}. \ce{La2H5} with $I4_1md$ symmetry has been proposed as a candidate in experiments performed by Conder et al.\cite{conder_large_1991} however we predict that the slightly different $I4_1/amd$ structure is more stable.

At high pressures above 10 GPa, a number of hydrides are found to compete, including \ce{LaH3} with $I4/mmm$ structure. Experiments by Laniel et al. found \ce{LaH3} (\fcc), $\ce{LaH_{\sim 4}}$ ($Cmcm$), $\ce{LaH_{4+\delta}}$ ($I4/mmm$), \ce{La4H23} ($Pm\overline{3}n$), $\ce{LaH_{6+\delta}}$ (\bcc{}), $\ce{LaH_{9+\delta}}$ (\hcp{}), and $\ce{LaH_{10+\delta}}$ (\fcc{}), mostly above 100 GPa \cite{laniel_high-pressure_2022}. The superconducting \ce{LaH10} appears in the phase diagram, unlike in previous predictions where Guan et al.\cite{guan_combining_2021} found \ce{LaH11} and \ce{LaH16} to be stable. \ce{LaH10} was found stable only when \ce{LaH11} was not considered. A more conclusive prediction in the relative stability of LaH10 and LaH11 would be made by inclusion of entropic and nuclear quantum effects, which is beyond the scope of this work. Only \ce{LaH10} was predicted originally and observed experimentally\cite{liu_potential_2017,geballe_synthesis_2018,somayazulu_evidence_2019}. 

\begin{figure}
\input{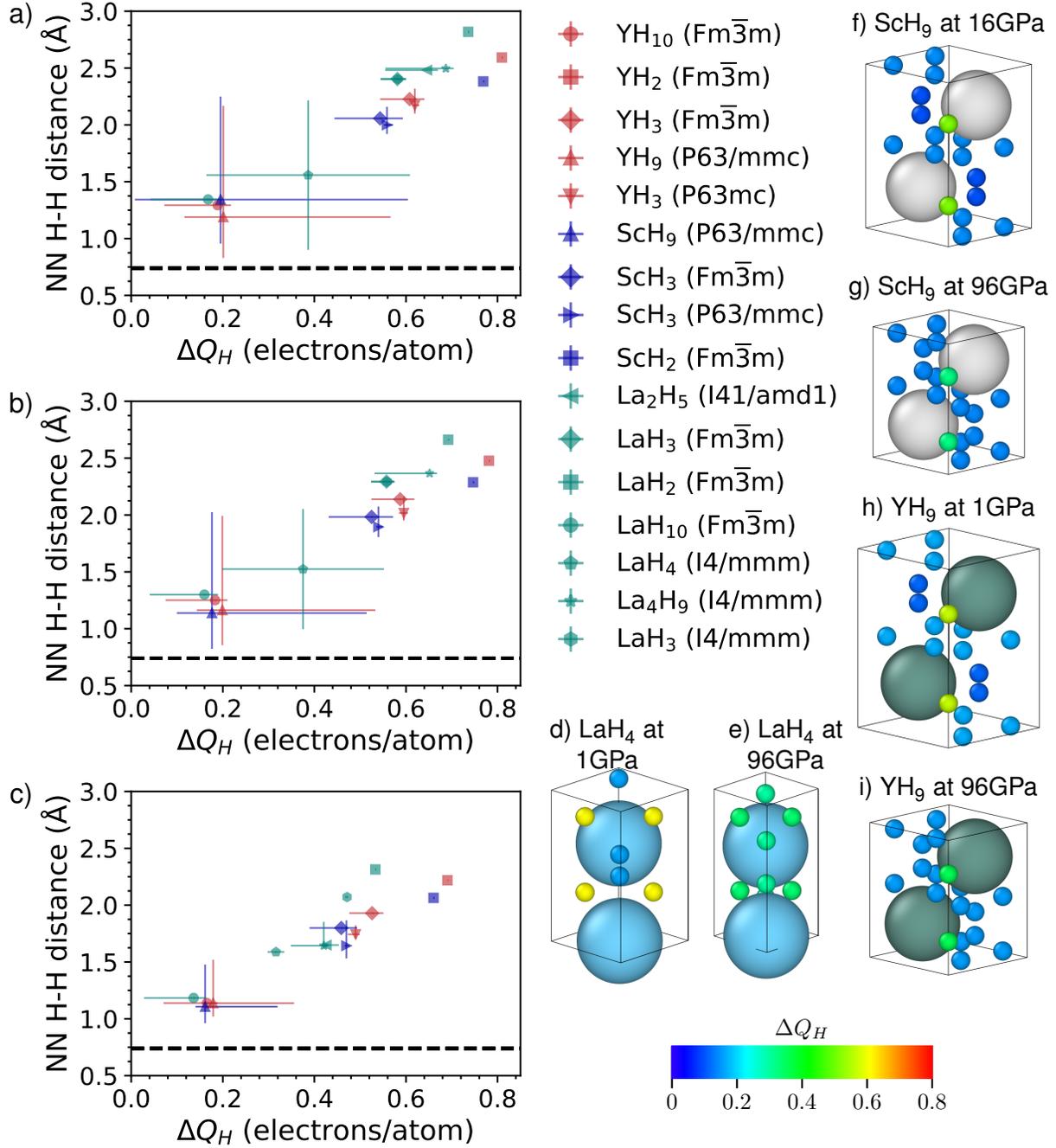}
\caption{Nearest neighbor distance between hydrogen atoms (\dnn{}) plotted against the change in the number of electrons gained per hydrogen atom (\qh{}) at a) 1 GPa, b) 16 GPa and c) 96 GPa. The plotted points are the average values and the errorbars are the range of values observed in the compound.}\label{fig:chg}
\end{figure}

To probe the local environments of hydrogen in the different hydrides, we show the nearest-neighbor H-H distance (\dnn{}) and number of electrons gained per hydrogen atom ($\Delta Q_H$) for each hydride that appears on the phase diagram at 1 GPa, 16 GPa and 96 GPa. General trends expected due the chemistry of hydrogen atoms hold. These include that for all the hydrides, the hydrogen atoms always gain electrons in all the hydrides, but never more than one. This is in contrast to hydrogen dissolved in electrolytes which typically lose their electron. Additionally for all hydrides, $\Delta Q_H$ taken up by hydrogen slightly decreases with increasing pressure as shown in \fref{fig:chg} and in more detail in the \siref{3}. Another expected trend is that the value of \dnn{} increases with increasing size of the metal ion for the same compound. 

As the hydrogen to metal ratio increases, the average $\Delta Q_H$ decreases as there are fewer available electrons for each hydrogen from the metal. A correlated trend is that the average \dnn{} decreases with increasing hydrogen to metal ratio. The closer \qh{} is to zero and the closer \dnn{} is to the molecular hydrogen bond length of 0.74\AA{}, the more the character of the hydrogen atoms is to that of molecular hydrogen. The error bars in \fref{fig:chg} show the range of values present for each particular compound and demonstrate that there is generally a wider variety of hydrogen environments with increasing hydrogen to metal ratio. We also show some color coded hydrogen atoms in some structures that demonstrate this in \fref{fig:chg}. At lower pressures in the \ce{MH9} structures and \ce{LaH4}, there are clearly pairs of hydrogen atoms with molecular character with very low $\Delta Q_H$ and bond length close to the hydrogen molecule distance. As pressure increases to 96 GPa, the hydrogen atoms lose this molecular character, gaining more charge and moving farther apart, despite the decrease in available volume in the unit cell. This change is desirable in the case of superconductivity as the character of the material approaches that of metallic hydrogen\cite{ashcroft_metallic_1968}. For hydrides with lower hydrogen content such as the \ce{MH2} (\fcc{}) compounds, there is only one type of hydrogen environment corresponding to the filling of octahedral sites in the metal sublattice. It is only the superhydrides with hydrogen to metal ratios greater than or equal to 4 that show hydrogen character very close to molecular hydrogen.

The differences within each compound in the hydrogen environments generally diminish with increasing pressure. This is especially true for the structures that contain motifs very similar to molecular hydrogen. We found in many cases that sometimes the DFT energies of these structures can be very close to competing stable hydrides even though they are clearly just lower hydrides suspended in a fluid of molecular hydrogen and would degas easily. Full consideration of entropic and nuclear quantum effects would more consistently resolve this source of error but is beyond the scope of this study. The departure from having molecular hydrogen is a sign of stabilization of the hydrogen within the hydride. Even as the available volume decreases with increases with pressure, the hydrogen atoms get farther apart.

In general for the same structure, \qh{} for the yttrium compounds is slightly higher than scandium by about 0.05 electrons per atom due to the corresponding ionization potentials of 6.217 eV and 6.561 eV respectively. It is therefore more favorable for yttrium to  donate electrons to the hydrogen atoms. Interestingly, for the same structure, yttrium consistently donates slightly fewer electrons than lanthanum despite lanthanum having an even lower ionization potential of 5.577 eV. The interplay between role of $f$ electrons, larger ionic radius and ionization potential is therefore not trivial.
 

\subsection{Period 3 Transition Metals}

\begin{figure*}[t!]
    \centering
    \begin{subfigure}[t]{0.6\textwidth}
        \centering
        \includegraphics[height=6cm]{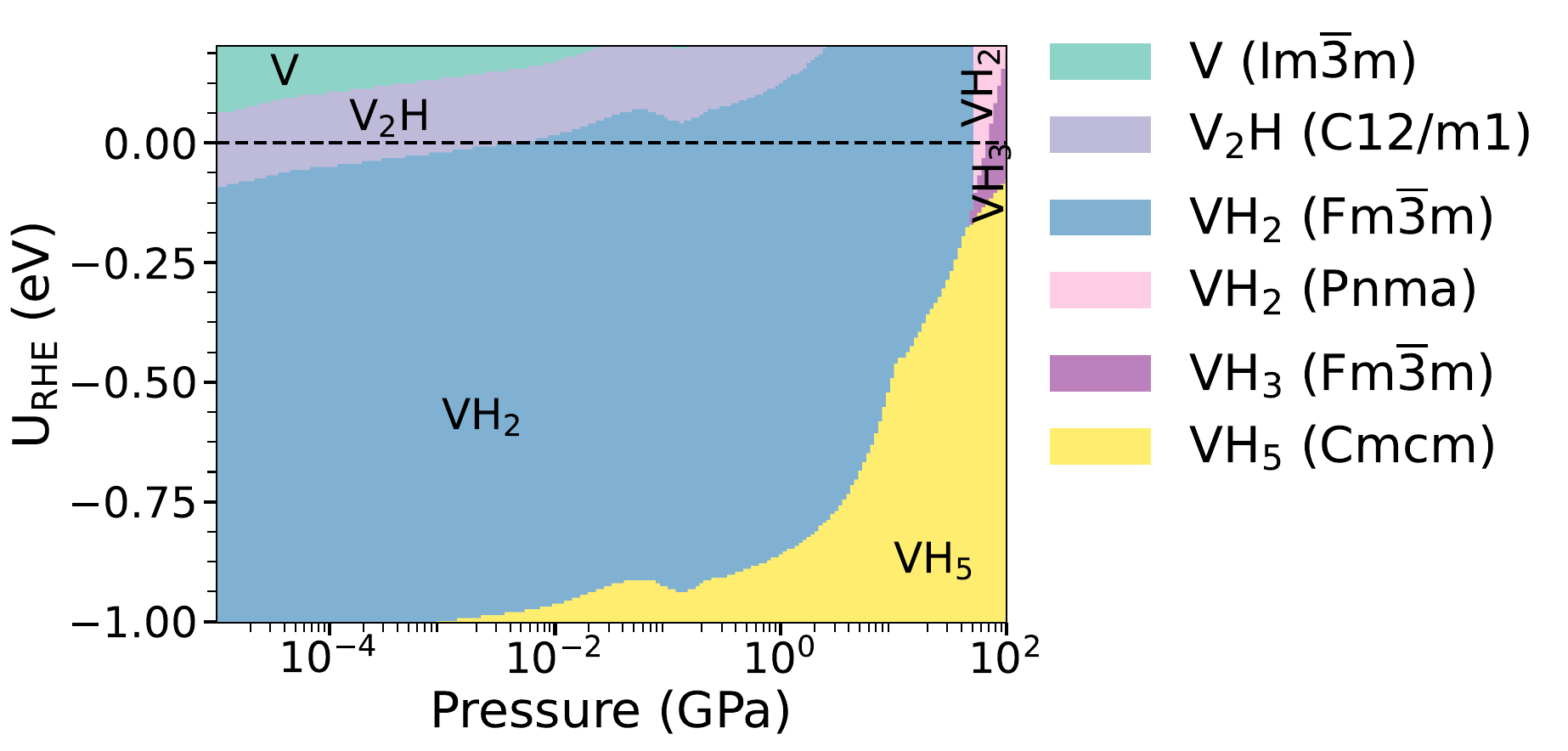}
        \caption{}\label{fig:vh}
    \end{subfigure}\\
    \begin{subfigure}[t]{0.6\textwidth}
        \centering
        \includegraphics[height=6cm]{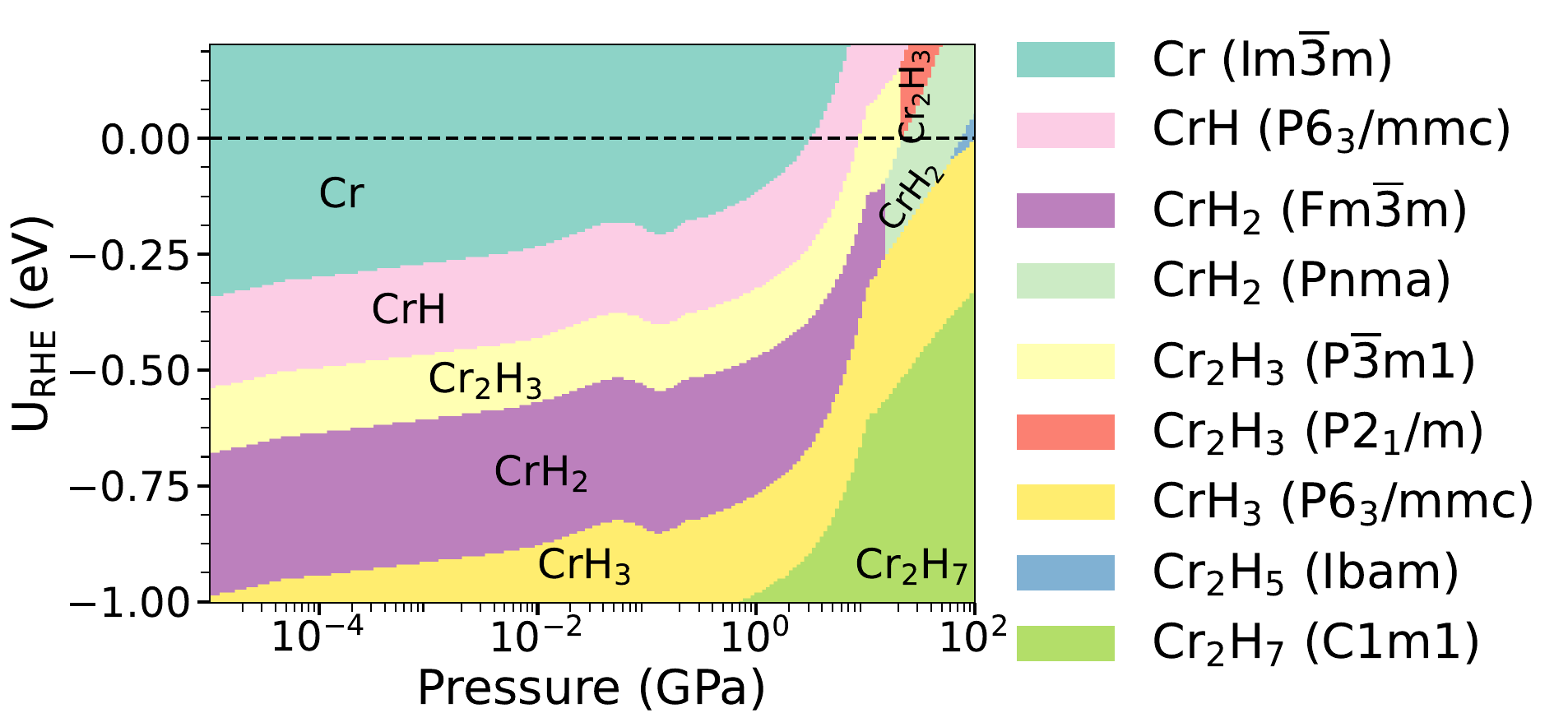}
        \caption{}\label{fig:crh}
    \end{subfigure}\\
    \caption{\pp{ phase diagrams for a) vanadium and b) chromium hydrides. The dashed line is a guide to the eye for $U_{\text{RHE}}=0$}}
\end{figure*}

\subsection{Vanadium Hydride}

Vanadium hydrides are considered for their excellent hydrogen storage capabilities \cite{kumar_development_2017} and also have applications as ammonia synthesis catalysts \cite{cao_vanadium_2021}. Vanadium hydrides have also been considered as cathodes for all-solid-state batteries \cite{matsumura_vanadium_2019} and flow batteries \cite{weng_high-voltage_2018}.

The predicted \pp{} phase diagram is shown in \fref{fig:vh} and shows that vanadium readily forms hydrides. The first hydride encountered is the \ce{V2H} ($C12/m1$) phase. This phase is known experimentally and is highly thermally stable which is the major limiting factor in its use for vanadium-based hydrogen storage \cite{kumar_development_2017}. On further hydrogen pressurization, the \ce{VH2} (\fcc{}) phase nucleates again in agreement with our prediction \cite{kumar_development_2017}. We predict the formation of other hydrides above 10 GPa, namely \ce{VH2} ($Pnma$), \ce{VH3} (\fcc{}) and \ce{VH5} however these have not been observed experimentally. There does not appear to be any accessible phase experimentally beyond \ce{VH2} below megabar pressures.

\subsection{Chromium Hydride}
The chromium-hydrogen phase diagram is known to be rich, containing gaseous phases \cite{gaydon_band_1937}, complexes \cite{wang_chromium_2003} and solid hydrides \cite{marizy_synthesis_2018}. Chromium has been known to form solid hydrides with hydrogen ratios greater than one since 1926 \cite{weichselfelder_uber_1926,marizy_synthesis_2018}. In fact, the formation of chromium hydride in the electroplating of chromium is unavoidable, making it an important intermediate to study for electrochemical processes.

In \fref{fig:crh} we show the predicted \pp{} diagram for the chromium-hydride system. Without pressure and increasingly negative $U_{\text{RHE}}$, we go from the pure \ce{Cr} (\bcc{}) to the HCP structure of \ce{CrH} (\hcp{}) with hydrogen taking the tetrahedral interstices. $\ce{CrH_{\sim 1}}$ with an HCP (\hcp) structure for the chromium sub-lattice is known to be formed during the electroplating of chromium \cite{khan_properties_1976}. Interestingly, \ce{CrH_{0.97}} is softer than \ce{Cr} \cite{khan_properties_1976} whereas hydrogen typically hardens metals due to hydrogen embrittlement \cite{wipf_metal-hydride_1997} which typically hinders electrochemical hydrogenation experiments due to crumbling of electrodes. We predict that a \ce{Cr2H3} compound with either $P\overline{3}m1$ symmetry at low pressure transitioning to $P2_1/m$ above 15 GPa, is formed. In 1947, Snavely et al. claimed they synthesized of a compound with \ce{CrH_{1.7}} stoichiometry and FCC structure\cite{snavely_theory_1947}. This result has never been experimentally reproduced however the \pp{} diagram suggests the existence of a similar stoichiometry albeit with $P\overline{3}m1$ structure below 11 GPa and a $P2_1/m$ structure above 11 GPa predicted computationally by Yu et al. \cite{yu_pressure-driven_2015}. With increasingly negative potential, a \ce{CrH2} (\fcc{}) compound stabilizes, which takes on a $Pnma$ structure above 10 GPa, then finally a \ce{CrH3} (\hcp{}) structure is predicted to be stable. 

Along the pressure axis the predicted phases go from \ce{Cr} (\bcc{}) to the \ce{CrH} (\hcp{}) phase around 1 GPa. Marizy et al. synthesized \ce{CrH} at 3 GPa in DAC experiments\cite{marizy_synthesis_2018}. The \ce{CrH} phase is predicted to be followed by the \ce{Cr2H3} phase with $P2_1/m$ symmetry at $\sim$15 GPa whereas in experiment this is found to occur at 19 GPa but to a different $C2/m$ structure after filling the tetrahedral sites of \ce{CrH} \cite{marizy_synthesis_2018}. Finally, \ce{CrH2} ($Pnma$) is predicted to form above $\sim$11 GPa, which was also synthesized by Marizy et al.\cite{marizy_synthesis_2018} above 31 GPa. We also predict the formation of a \ce{Cr2H5} compound with $Ibam$ symmetry above 80 GPa. At $U_{\text{RHE}}$ below $-0.25$eV and pressure above 1 GPa, we predict the potential formation of \ce{Cr2H7} (P21/m). For the conditions considered in this work, this phase can only be stabilized under conditions of combined pressure and potential.

The \pp{} diagram for chromium hydrides agrees quite well with experiment for both the purely electrochemical and pressure loading cases. Using both approaches simultaneously suggests that \ce{Cr2H3} ($I4/mmm$) and \ce{CrH2} (\fcc{}) can be formed at gigapascal pressures. These phases would be difficult to access only electrochemically due to the HER at standard pressure.

\section{Conclusions}
For the metal hydride systems considered, hydrogen is taken up more easily i.e. at less negative $U_{\text{RHE}}$ and at lower pressures going down the periodic table. For example, the transition from the \ce{MH2} phase to the \ce{MH3} phase occurs at -0.3 eV, -0.1 eV, and 0.0 eV for scandium, yttrium and lanthanum respectively at $10^{-5}$GPa. Note that the lanthanum transition is predicted by disregarding the \ce{La2H5} and \ce{La4H9} phases as shown in \siref{2}. At $U_{\text{RHE}}=0$ with varying pressure, the same transition happens at $\sim$3 GPa, $\sim$10$^{-4}$GPa for scandium and yttrium respectively while \ce{LaH3} is readily formed at ambient pressure. At any combination of $U_{\text{RHE}}$ and pressure, the ratio of hydrogen in lanthanum is higher than yttrium which in turn is higher than scandium. The variety of structures also increases leading to a richer phase diagram in lanthanum. 
The observed trend can be correlated with decreasing metal ionization potential and increasing ionic radius. The ionization potential plays a part in splitting the hydrogen atoms, facilitating the formation of atomic-like hydrogen with longer \ce{H-H} bonds by donating charge as demonstrated by Bader charge analysis.  The metal ion radius plays a role in allowing more hydrogen atoms to be packed around the ion, especially in the cage-like clathrate structures that form in extreme-pressure conditions\cite{boeri_road_2019}.

The \pp{} approach offers new pathways to hydride synthesis and the phase diagrams in this work demonstrate the current predictive ability of the methodology. The structures considered in this work largely came from crystal structure search data and the resulting phase diagrams agree quite well with experiment. As more structure data becomes available and effects of finite temperature and nuclear quantum effects are included, we expect the predictions to improve. Considering non-stoichiometry of the hydrides can also be important, particularly at low to moderate pressures\cite{guan_thermodynamic_2025}. The challenge now is to develop experiments to test predictions under combined pressure and potential, particularly above 1 GPa. There also remains the question of how long different hydrides remain stable after pressure and potential are released. This may be sufficiently long for some applications and characterization techniques. Interestingly, Kataoka\cite{kataoka_face-centered-cubic_2021} were able to synthesize \ce{YH3} (\fcc{}) at ambient pressures using ball milling techniques and even showed that it maintained its structure without maintained external pressure emphasizing the potential of the chosen pathway in the hydride synthesis.


\begin{acknowledgement}

MKP and VV authors thank Google LLC. RJH acknowledges the support of the National Science Foundation (DMR-2104881) and DOE-NNSA (DE-NA0004153). The authors also thank Karen Sugano, Florian Meltzer, Yet Ming Chiang and Eva Zurek for discussion and feedback.

\end{acknowledgement}

\begin{suppinfo}

Supporting data and figures are given in the Supporting Information

\end{suppinfo}

\bibliography{achemso-demo}

\renewcommand{\thefigure}{S\arabic{figure}}
\setcounter{figure}{0}

\begin{center}
    \section{Supporting Information for: Stability and Structure of Binary Metal Hydrides under Pressure, Electrochemical Potential and Combined Pressure-Electrochemistry}
\end{center}

\maketitle

\begin{figure}
    \centering
    \includegraphics[width=12cm]{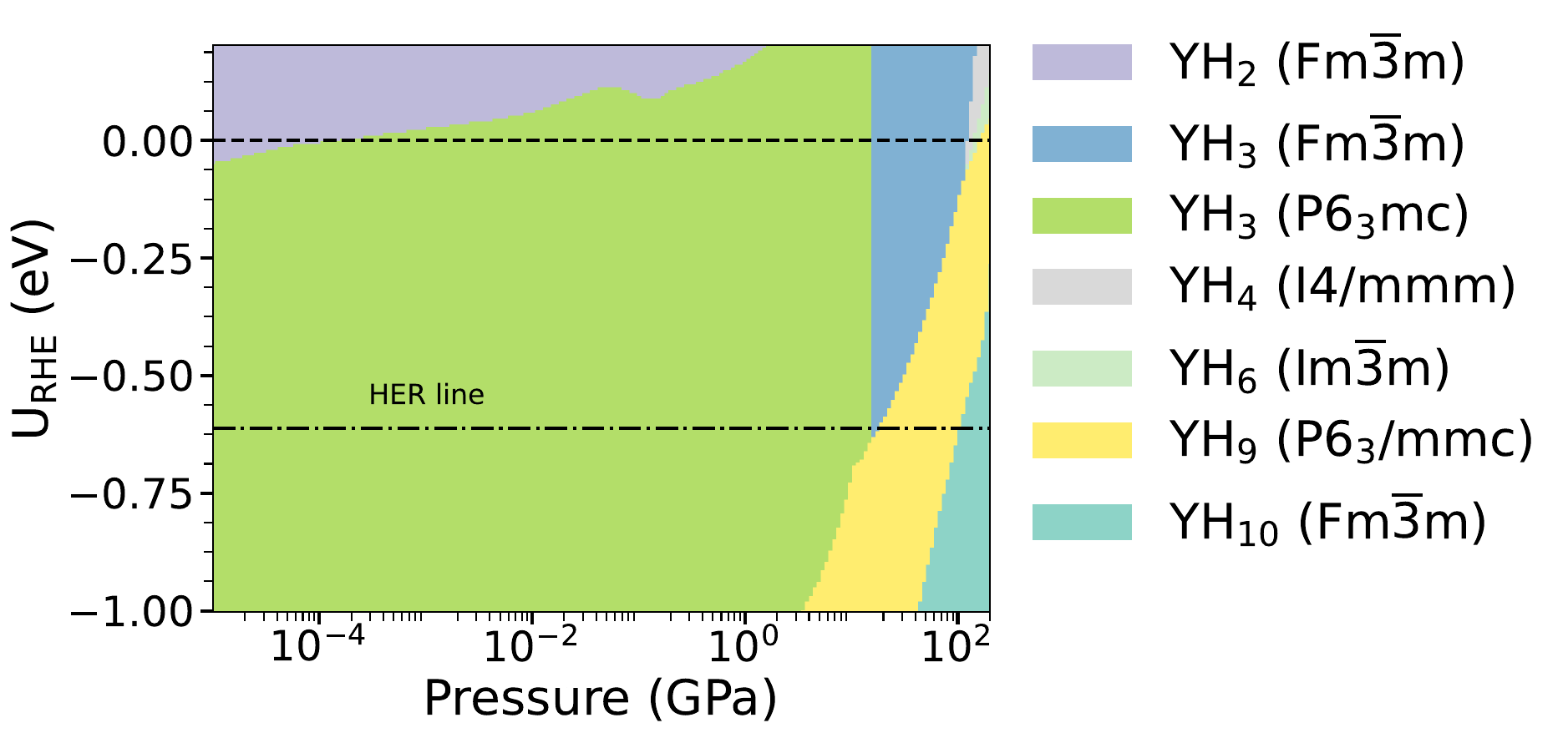}
    \caption{\pp{} diagram for Y-H system up to 200GPa. The experimentally synthesized \ce{YH4} amd \ce{YH6} are found stable.}
\end{figure}

\begin{figure}
    \centering
    \includegraphics[width=12cm]{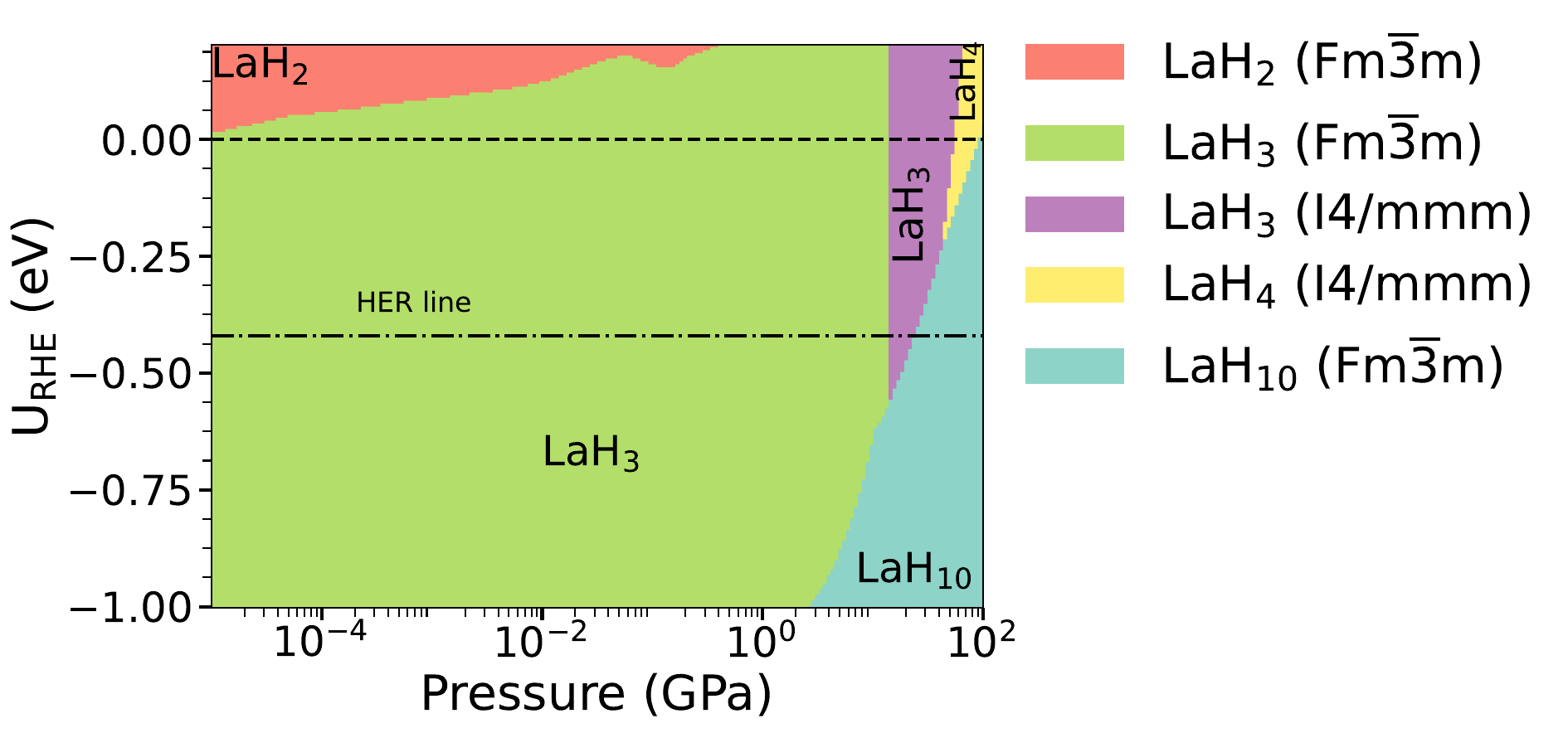}
    \caption{\pp{} diagram for La-H but without considering the \ce{La2H5} phase.}
\end{figure}

\begin{figure}
     \centering
     \subfloat[width=4cm][\ce{YH2}]{\includegraphics[width=6cm]{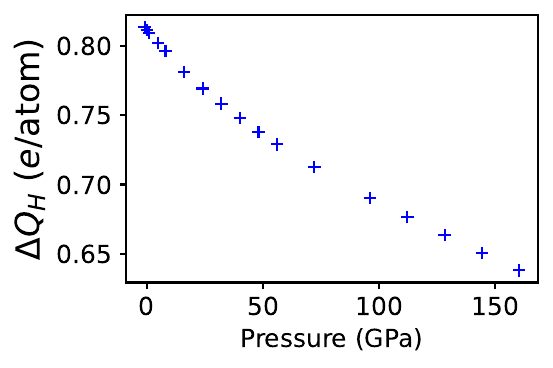}\label{qvsp_yh2}}
     \subfloat[width=4cm][\ce{YH3}(\hcp{})]{\includegraphics[width=6cm]{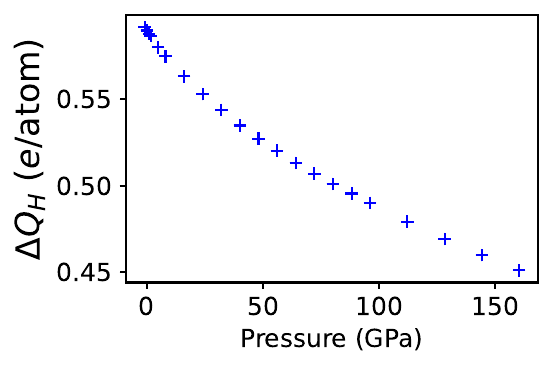}\label{qvsp_yh3}}\\
     \subfloat[width=4cm][\ce{YH3}(\fcc{})]{\includegraphics[width=6cm]{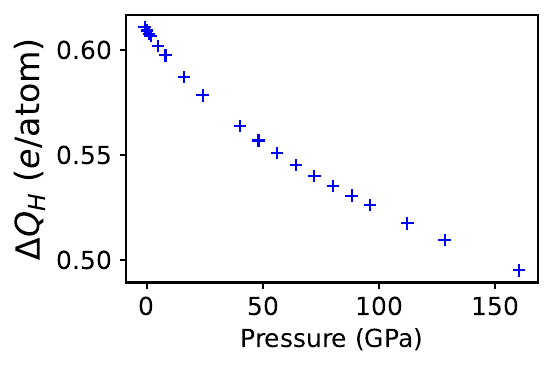}\label{qvsp_yh3fcc}}
     \subfloat[width=4cm][\ce{YH4}]{\includegraphics[width=6cm]{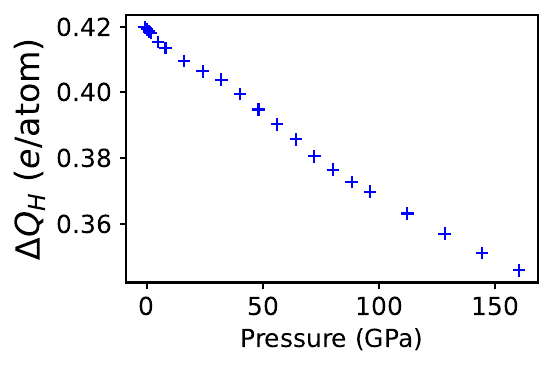}\label{qvsp_yh4}}\\
     \subfloat[width=4cm][\ce{YH6}]{\includegraphics[width=6cm]{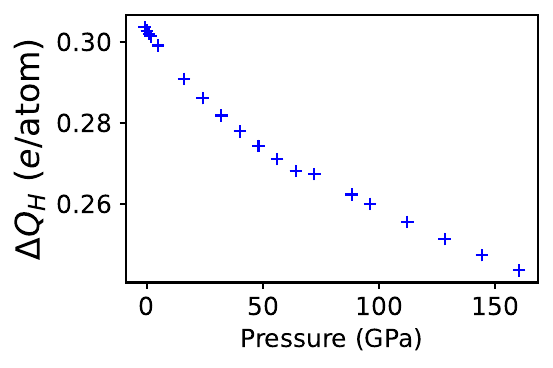}\label{qvsp_yh6}}
     \subfloat[width=4cm][\ce{YH9}]{\includegraphics[width=6cm]{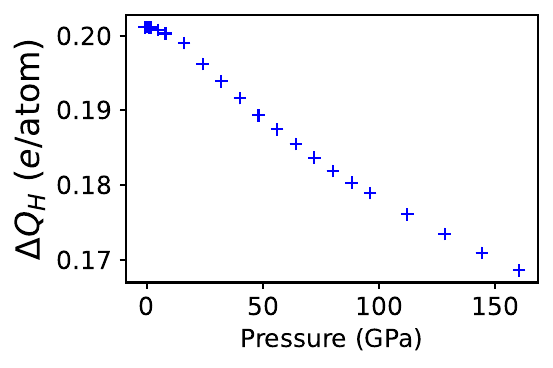}\label{qvsp_yh9}}\\
     \subfloat[width=4cm][\ce{YH10}]{\includegraphics[width=6cm]{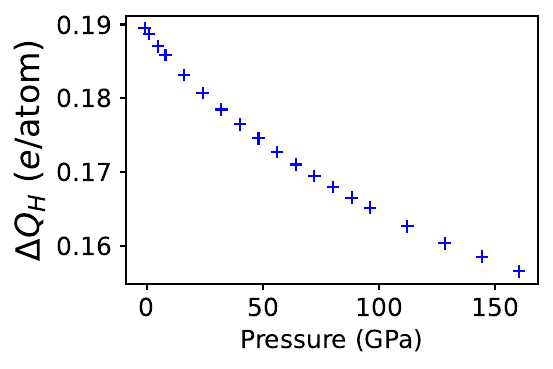}\label{qvsp_yh10}}\\
     \caption{Average $H$ charge gained for yttrium hydrides appearing on the phase diagram. The change over the pressure range considered is typically $\sim$0.15$e$/$H$ except for \ce{YH10} where it is only $\sim$0.4$e$/$H$. Note that the value of these changes depends on the charge partitioning scheme, the variation under different conditions is the significant insight. Despite the decreasing volume as pressure increases there is clearly an increased tendency to transfer electrons to hydrogen as pressure increases in all cases.}
     \label{steady_state}
\end{figure}


\end{document}